\documentclass[%
 aip,
 amsmath,amssymb,
 reprint,%
]{revtex4-1}

\usepackage{graphicx}% Include figure files
\usepackage{dcolumn}% Align table columns on decimal point
\usepackage{bm}% bold math
\usepackage[utf8]{inputenc}
\usepackage[T1]{fontenc}
\usepackage{mathptmx}
\usepackage{etoolbox}

 \usepackage{comment}
 \usepackage{siunitx}
  \usepackage{tikz} 
  \usepackage{soul}
  \usepackage[version=3]{mhchem}

\makeatletter
\def\@email#1#2{%
 \endgroup
 \patchcmd{\titleblock@produce}
  {\frontmatter@RRAPformat}
  {\frontmatter@RRAPformat{\produce@RRAP{*#1\href{mailto:#2}{#2}}}\frontmatter@RRAPformat}
  {}{}
}%
\makeatother
\begin{document}

\preprint{}

\title[\small{}]{Ultrafast Electron Temperature Dynamics in Spintronic Terahertz Emitters Studied by Optical-Pump Terahertz-Probe spectroscopy}

\author{Felix Selz*}
  \email{felix.selz@physik.uni-freiburg.de}
\affiliation{ 
Fraunhofer Institute for Industrial Mathematics ITWM, Department of Materials Characterization and Testing, 67663 Kaiserslautern, Germany
}%
\affiliation{School of Engineering, Brown University, Providence, Rhode Island 02912, USA}
\affiliation{Department of Physics and Research Center OPTIMAS, RPTU Kaiserslautern-Landau, 67663 Kaiserslautern, Germany}
\affiliation{Institute of Physics, University of Freiburg, 79100 Freiburg, Germany}

\author{Johanna Kölbel}%
\affiliation{School of Engineering, Brown University, Providence, Rhode Island 02912, USA}

\author{Felix Paries}
\affiliation{ 
Fraunhofer Institute for Industrial Mathematics ITWM, Department of Materials Characterization and Testing, 67663 Kaiserslautern, Germany
}%
\affiliation{Department of Physics and Research Center OPTIMAS, RPTU Kaiserslautern-Landau, 67663 Kaiserslautern, Germany}

\author{Georg von Freymann}
\affiliation{ 
Fraunhofer Institute for Industrial Mathematics ITWM, Department of Materials Characterization and Testing, 67663 Kaiserslautern, Germany
}%
\affiliation{Department of Physics and Research Center OPTIMAS, RPTU Kaiserslautern-Landau, 67663 Kaiserslautern, Germany}

\author{Daniel Molter}
\affiliation{ 
Fraunhofer Institute for Industrial Mathematics ITWM, Department of Materials Characterization and Testing, 67663 Kaiserslautern, Germany
}%

\author{Daniel M. Mittleman}
\affiliation{School of Engineering, Brown University, Providence, Rhode Island 02912, USA}

\date{\today}% It is always \today, today,

\begin{abstract}
Spintronic terahertz emitters (STEs) pumped by femtosecond lasers have become a widely used source of broadband terahertz radiation. However, the strength of the emitted field is limited in part by the optical damage threshold at the pump wavelength. Thermal management of STEs can be improved by understanding electron temperature relaxation in the spintronic metal layer. Here, we present a measurement of electron temperature dynamics on a picosecond timescale using optical-pump terahertz-probe spectroscopy. We observe that the optical pump induces a change in terahertz transmission through the STE. By analyzing the resulting signal with a two-temperature model, we extract the dynamic electron temperature of the STE. This approach offers an advantage over other methods by avoiding additional heating of the sample by the probe pulse, making it particularly suitable for studying cumulative heating effects, which are believed to contribute to optical damage under MHz repetition rate pumping.
\end{abstract}

\maketitle
\section*{Introduction}
Designing efficient emission and detection devices in the terahertz regime ($\SI{0.1}{THz}$ to $\SI{10}{THz}$, between microwave electronics and infrared optics) has been a long-standing challenge. 
However, significant progress has been made in recent years, for example with the development of spintronic terahertz emitters (STEs) that offer much higher bandwidth than other commonly used emitters such as photoconductive antennas~\cite{kampfrath_terahertz_2013, seifert_spintronic_2022, torosyan_optimized_2018, ilyakov_terahertz-wave_2022, papaioannou_efficient_2018, nandi_antenna-coupled_2019, ilyakov_efficient_2023, seifert_efficient_2016}. STEs rely on the current-to-field conversion of a photocurrent originating from a femtosecond laser pulse excited spin current and spin-to-charge conversion in a magnetic heterostructure~\cite{seifert_spintronic_2022}. They yield the advantage of a simple metallic thin-film-stack design which can be sputtered onto various substrates and therefore be fabricated and scaled at low cost~\cite{torosyan_optimized_2018}. Also, STEs can be pumped over a broad frequency range~\cite{ilyakov_terahertz-wave_2022,papaioannou_efficient_2018,nandi_antenna-coupled_2019,ilyakov_efficient_2023}, which enables a very adaptable implementation in table-top terahertz spectroscopy setups. 

For sufficiently strong pump lasers, the limiting factor of the terahertz output power is the optical damage threshold. We have recently shown that the optical damage threshold of fiber-tip spintronic terahertz emitters can be classified into two regimes corresponding to two different destruction mechanisms~\cite{paries_optical_2024}.

When an optical laser pulse hits the STE, it raises both the electron temperature in the metal layers and the lattice temperature. Within a few tens of picoseconds, the electron and phonon systems equilibrate to a new steady-state temperature, which is higher than the initial equilibrium temperature before the pulse. This steady-state temperature relaxes much more slowly -- on the order of microseconds to milliseconds -- due to thermalization with the environment trough coupling to the substrate~\cite{agarwal_ultrafast_2021}. As a result, when the time between successive laser pulses is short (e.g., at repetition rates above {\SI{4}{MHz}}), heat can accumulate in the system. This can lead to a progressive increase in the STE’s steady-state temperature over time. This gradual increase in steady-state temperature drives atomic interlayer diffusion, which is the dominant degradation mechanism of the STE at high repetition rates~\cite{paries_optical_2024}. Other studies already targeted this heat driven destruction and introduced new methods for an improved thermal management of the STE~\cite{vogel_average_2022,vaitsi_rotating_2024,gandubert_spatial_2024}. Vogel \textit{et al.} employed a water cooled mount for the STE, improving the heat dissipation from the optical pump spot\cite{vogel_average_2022}. In a different approach, Vaitsi \textit{et al.} used a rotating STE mount, allowing to distribute the optical pump power over a larger area and thereby increasing the damage threshold\cite{vaitsi_rotating_2024}. In another study, Gandubert \textit{et~al.} showed, how optimizing the spatial distribution and temporal spreading of the optical pump pulses can enhance the STEs overall efficiency before reaching a destruction threshold.\cite{gandubert_spatial_2024}

To optimize thermal management in such cases, it is essential to understand the dynamics of both the electron and phonon systems in the STE. This includes, most importantly, the evolution of the electron temperature in the first few picoseconds following excitation, the electron–phonon coupling over the subsequent tens of picoseconds, and the longer-timescale heat transport from the pump spot to the surrounding area -- both laterally within the STE layer and vertically into the underlying substrate. Previous studies measured the electron temperature using techniques like photoemission spectroscopy~\cite{fann_direct_1992}, electron diffraction~\cite{volkov_investigation_1976}, Anti-Stokes photoluminescence spectroscopy~\cite{jollans_effective_2020}, or optical pump-probe spectroscopy~\cite{hayashi_determination_2022, hohlfeld2000electron, agarwal_ultrafast_2021}.

The electron temperature in STEs consisting of a tri-layer structure of W(3\,nm) | NiFe(3\,nm) | Pt(3\,nm) has previously been measured in a reflection geometry using optical-pump optical-probe spectroscopy for different pump fluences around {\SI{1}{mJ/cm^2}}. An increase in electron temperature above {\SI{1000}{K}} and relaxation dynamics on the order of several hundreds of ps back to a steady-state temperature were shown.{~\cite{agarwal_ultrafast_2021}} However, in such measurements, it is unclear how much the electron temperature is influenced by the probe. Such effects are particularly confounding in the case of accumulated thermal effects, where the probed region does not return to equilibrium between consecutive pump pulses, as also mentioned by the authors~\cite{agarwal_ultrafast_2021}. 
Here we present a different approach for investigating the electron temperature dynamics using an optical-pump terahertz-probe setup~{\cite{beard_terahertz_2002}} and a STE with a different ferromagnetic layer ($\mathrm{Fe}_{60}\mathrm{Co}_{20}\mathrm{B}_{20}$). 
Due to the low photon energy of the probe pulse, it does not induce a significant change in the electron temperature of the STE, thereby eliminating a potential source of ambiguity in the measurement. Furthermore, the experiments are conducted at a low repetition rate of 1\,kHz to enable an undisturbed observation of the electron temperature dynamics following the absorption of a single optical pump pulse. This approach ensures that cumulative heating effects from subsequent pulses, as discussed above, are avoided.
While the electron–phonon dynamics observed here occur on picosecond timescales where optical probe-induced perturbations are typically small, due to the low photon energy of the probe pulse, nonlinear absorption processes such as two- or multi-photon absorption are strongly suppressed\cite{boyd2008nonlinear}. This provides a minimally invasive probing approach that may be especially important in future applications involving ultrafast dynamics, fragile materials, or resonant optical transitions. The estimated terahertz probe field strengths used here are approximately 100\,kV/cm, which corresponds to a pulse energy of $E_\text{p, THz} \approx \SI{0.65}{\micro J}$.

\section*{Methods}
\begin{figure}[t]
    \centering
    \includegraphics[width=0.9\columnwidth]{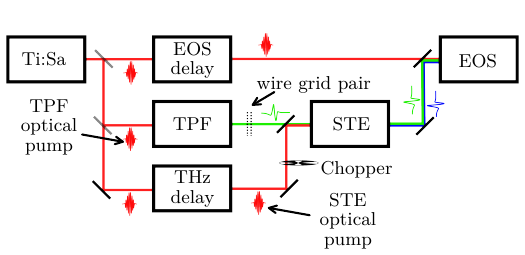}
    \caption{Schematic of the experimental setup. Optical beams in red (optical detection beam, TPF optical pump and STE optical pump), terahertz beams in green (TPF, terahertz probe beam) and blue (STE emission). Important components as the chopper for lock-in detection and the wire grid pair for polarization and intensity control of the terahertz probe pulse are displayed. A more detailed visualization of the temporal dynamics between the different pulses is shown in Figure 2. Ti:Sa - Titanium Sapphire laser; EOS - Electro Optical Sampling; TPF - Tilted Pulse Front terahertz generation; STE - Spintronic Terahertz Emitter.}
    \label{fig:Exp_setup}
\end{figure}

The experimental setup is described in detail in Ref.~\citenum{Selz_25} and a schematic version of the setup can be found in Figure \ref{fig:Exp_setup}. Briefly, three beam paths originate from an amplified Ti:sapphire laser with a center wavelength of \SI{800}{nm}, a pulse length of \SI{80}{fs}, a repetition rate of \SI{1}{kHz} and a maximum output power of \SI{6}{W}. 
The STE is pumped by an optical beam at 800\,nm, using approximately 10\,\% of the laser output power. With a pulse energy of 0.1\,mJ and a pump spot size of 0.1\,cm$^2$, this corresponds to a fluence of 1\,mJ/cm$^2$ at the STE. The terahertz probe beam is generated via optical rectification in a {\ce{LiNbO3}} crystal using a tilted-pulse-front (TPF) setup, which allows for phase-matched generation of intense, single-cycle terahertz pulses{~\cite{Hebling_08}}. This process utilizes approximately 80\,\% of the laser power. The terahertz probe is focused onto the same position on the sample using parabolic mirrors, resulting in a probe spot size of approximately 0.05\,cm$^2$. These experimental parameters are kept constant for all experiments described here.
The optical pump beam is blocked after the STE, and the two terahertz signals corresponding to the probe and the terahertz beam generated by the STE are detected with electro-optic sampling and balanced detection~\cite{koch_terahertz_2023}, using the remaining $\sim$\SI{10}{\%} of the laser power. Delay stages control the relative time delay between the arrival of the terahertz probe pulse and the STE pulse at the detector. Pairs of wire grid polarizers before and after the sample position are used to further control the polarization and intensity of the different beams.

The STEs are manufactured by depositing a W(\SI{2.0}{nm})$\big\vert$$\mathrm{Fe}_{60}\mathrm{Co}_{20}\mathrm{B}_{20}$(\SI{1.8}{nm})$\big\vert$Pt(\SI{2.0}{nm}) tri-layer structure onto \SI{0.5}{mm} thick sapphire substrates employing RF-diode sputtering. Details about the manufacturing process can be found in Ref.~\citenum{paries_fiber-tip_2023}.
Magnets are mounted around the STE and are used to control the polarization of the terahertz pulses emitted by the STE~\cite{seifert_spintronic_2022}.

\section*{Results and discussion}
\begin{figure}[t]
    \centering
    \includegraphics[width=0.9\columnwidth]{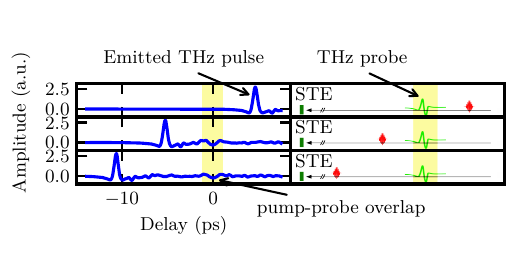}
    \caption{Measured terahertz waveform emitted from the STE (left panels, blue), with polarization parallel to the detection axis. Because the terahertz probe beam is not modulated by a chopper, the measured waveform originates solely from the emission of the STE (induced by the optical pump pulse). The optical pump pulse has a fluence of 1\,mJ/cm$^2$  (spot size: 0.1\,cm$^2$) and the terahertz probe pulse has a fluence of 0.013\,mJ/cm$^2$ (spot size: 0.05\,cm$^2$). The pump-probe delay is varied from top to bottom. The corresponding schematic on the right shows the relative timing of the optical pump pulse (red) and the terahertz probe pulse (green). When the optical pump follows the terahertz probe pulse, no additional feature appears. When the pump precedes the probe, a distinct peak emerges at a constant delay. The temporal overlap between the emitted terahertz pulse and the terahertz probe pulse is marked as pump-probe overlap.}
    \label{fig:first_obs_wavefollowing}
\end{figure}

During measurements, the chopper is placed in the optical pump path. Because the terahertz probe beam is not modulated by a chopper, the corresponding signal is not amplified, and only the emission from the STE (induced by the pump pulse) is measured.
Figure~\ref{fig:first_obs_wavefollowing} shows typical measurement results, as well as a schematic drawing of the different relative time delays. In the top panel, the optical pump pulse  reaches the STE wafer \textit{after} the terahertz probe pulse; in the middle and bottom panel, the optical pump pulse reaches the STE wafer \textit{before} the terahertz probe pulse with a different relative time delay between them. In the latter two cases, an additional peak is visible in the measured waveforms at the time delay of the terahertz probe pulse (highlighted in yellow). This indicates that the presence of the optical pump pulse modifies the signal strength of the terahertz probe pulse at the detector. To further investigate this effect, we rotate the polarization of the infrared detection beam with respect to the z-axis of the ZnTe detection crystal to suppress the STE signal contribution~{\cite{planken2001measurement}}. In this configuration, the chopper is still placed in the optical STE pump path, and the lock-in amplifier is referenced to that modulation. The measured signal therefore corresponds to optical STE pump-induced changes in the transmitted terahertz probe pulse.

\begin{figure}[t]
    \centering
    \includegraphics[width=0.9\columnwidth]{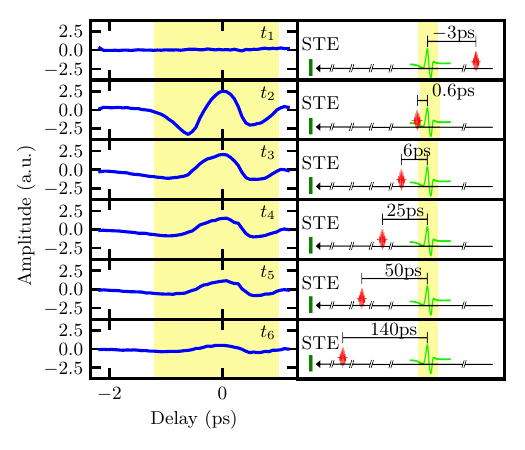}
    \caption{STE terahertz waveform with polarization perpendicular to the detection polarization. All other experimental parameters are the same as described in the caption of Figure~2.
    The optical pump pulse delay is changed from top to bottom and schematically drawn on the right side. When the optical pump pulse follows the terahertz probe pulse, no additional peak is observed in the waveform at the terahertz probe pulse delay. When the optical pump pulse precedes the terahertz probe pulse, an additional peak at a constant delay is observed, with decreasing amplitude when the time delay between optical pump pulse and terahertz probe pulse is increased. The time delay of the terahertz probe pulse is highlighted in both the measurement and the schematic drawing.}
    \label{fig:wavefollowing_low_signal}
\end{figure}

Figure~\ref{fig:wavefollowing_low_signal} shows a close-up view of the additional peak in the measured signal, along with a schematic representation of the different relative timing configurations of the optical pump pulses.
When the optical pump pulse arrives after the terahertz probe pulse, shown in the top panel, no peak is visible.
However, when the pump pulse excites the wafer before the terahertz probe pulse arrives (the second and all lower panels in Fig.~\ref{fig:wavefollowing_low_signal}), the additional peak appears in the time trace of the terahertz waveform, always at the same time delay as the terahertz probe pulse.

\begin{figure}[t]
    \centering
    \includegraphics[width=0.9\columnwidth]{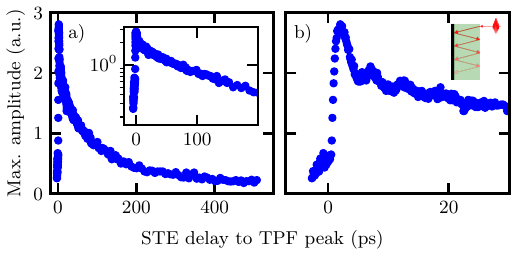}
    \caption{a) Maximum amplitude of the additional peak in the STE terahertz waveform induced by the terahertz probe pulse in the experiment shown in Figure 3. The inset shows the same plot on a logarithmic scale. b) A zoomed-in view of the first \SI{30}{ps} of the same decay as in (a). Here, multiple oscillations in the first \SI{20}{ps} are observed. The oscillations can be explained by multiple reflections in the STE wafer as indicated in the inset.}
    \label{fig:waveform_decay_zoom}
\end{figure}

Following the panels from top to bottom, it is clear that the amplitude of the additional observable peak decreases with increasing time delay between the optical pump pulse and the terahertz probe pulse. This signal must originate from the influence of the optical pump pulse on the sample. The optical pump pulse, in addition to generating a terahertz signal, also produces an instantaneous increase of the electron temperature, which then relaxes on a picosecond time scale.\cite{seifert_efficient_2016, wu2021principles, vaitsi_rotating_2024}. The optical pump pulse therefore varies the transmittivity of the terahertz probe pulse through the STE. Since this change is modulated at the frequency of the optical chopper in the optical STE pump path, it is detected in the lock-in measurement of the terahertz waveform.
Hayashi \textit{et al.}{\cite{hayashi_determination_2022}} demonstrated that changes in the optical reflectivity of metallic samples are sensitive to both electron and phonon temperatures, through modifications of the complex dielectric function. While an approximate proportionality between electron temperature and reflectivity can hold under specific conditions, the relationship is in general nontrivial and model-dependent.{\cite{hayashi_determination_2022}}  Since reflectivity and absorption both derive from the complex refractive index, they are intrinsically linked by the Kramers-Kronig relations. In our measurements, reflection is not negligible, and the observed change in transmission thus reflects a combined change in absorption and reflectivity.
Thus, by measuring the change in transmission, we can track the variation in electron temperature induced by the optical pump. The amplitude of the observed additional peak as a function of the relative time delay between the optical pump pulse and the terahertz probe pulse is plotted in Figure~\ref{fig:waveform_decay_zoom}a and~\ref{fig:waveform_decay_zoom}b.

 % the amplitude of the observed additional peak is plotted as a function of the relative time delay between the optical pump pulse and the terahertz probe pulse. 
In Figure~\ref{fig:waveform_decay_zoom}a, we observe a rapid initial drop, followed by a slower exponential decay with a time constant of $\sim$\SI{50}{ps}. The exponential decay after the rapid initial drop is illustrated in the logarithmic plot in the inset of figure~\ref{fig:waveform_decay_zoom}a where a linear behavior is visible.
In the first \SI{30}{ps} (Figure~4b), we observe oscillations on a timescale of $\sim$\SI{5}{ps}, superimposed on the slow decay. The origin of these oscillations are multi-pulse excitations, as a result of multiple reflections in the sapphire substrate of the STE. Given the thickness of the sapphire substrate of \SI{0.5}{mm} and the refractive index of sapphire~\cite{thomas_frequency_1998} at \SI{800}{nm}, $n_\text{sapphire} = 1.7$, the time delay of a pulse traveling through the sapphire substrate for one round trip due to reflections is $t=\frac{2\cdot\SI{0.5}{mm}}{c/n_\text{sapphire}}\approx\SI{5.67}{ps}$. This matches very well with the observed dynamics of the oscillations. It is therefore likely that these oscillations originate from multiple reflections in the sapphire substrate, as illustrated in the inset in Figure~4b.

To understand the different timescales of the observed change in transmission, we consider the sequence of interactions occurring in magnetic materials after excitation by a femtosecond laser pulse. First, the energy of the optical pulse is absorbed by electrons, forcing the system out of equilibrium~{\cite{bigot_coherent_2009}}. The electrons rapidly relax into hot thermalized populations on timescales below 100~fs~{\cite{bigot_coherent_2009}}. These dynamics are faster than the duration of the probing terahertz pulse and are therefore not resolved in our measurements. However, they are responsible for the abrupt increase in amplitude seen in Figure~4b at pump–probe delays shorter than 1\,ps. Subsequently, electron–phonon coupling becomes dominant. This interaction occurs over a few picoseconds and gives rise to the fast decay observed in Figure~4b within the first 5~ps after excitation. At this point, the electron and phonon systems have reached thermal equilibrium.
The subsequent dynamics involve heat dissipation from the pump spot into the surrounding system.

This process involves two parallel channels: lateral heat transport within the STE layer, and vertical heat transfer across the STE–sapphire interface, which is limited by electron–interface scattering~{\cite{agarwal_ultrafast_2021, guo_heat_2011}}. Although both processes occur simultaneously, they operate on different timescales.
The in-plane heat diffusion is relatively fast on the timescale of a few tens of ps, while the interfacial heat transfer is slower due to thermal boundary resistance and on the timescale of a few hundreds of ps~{\cite{agarwal_ultrafast_2021}}. Together, these mechanisms contribute to the slower decay observed in Figure~4b, which extends to approximately 400~ps after excitation.
    
We note that our experimental time resolution does not allow us to capture the ultrafast rise of the electron temperature in the first tens of femtoseconds following optical excitation. As such, the peak electron temperature prior to electron–phonon equilibration is not directly accessible in our measurements. Our analysis is therefore limited to the evolution of the electron–lattice system after thermalization, and the estimated temperature values reflect this regime.

A quantitative description of these dynamics can be extracted from a simple two-temperature-model (TTM)~{\cite{anisimov_electron_1974, kaganov_relaxation_1957} as discussed in Ref.~\citenum{agarwal_ultrafast_2021} and Ref.~\citenum{atxitia_evidence_2010}. The TTM describes the temperatures of the electronic system and the lattice, $T_e$ and $T_l$, in a metal after an excitation of the electronic system. In our case, the excitation is the absorbed femtosecond pulse and the metal is the thin STE layer. In metallic thin films with thicknesses below the electron inelastic mean free path (IMFP), such as the ferromagnetic layers used in STEs, lateral transport of hot electrons is significantly suppressed compared to a bulk metal. As a result, energy relaxation immediately following optical excitation is not solely dominated by hot electron diffusion, but also electron–phonon coupling plays a critical role in temperature relaxation.~\cite{hohlfeld2000electron,giri2020review} After excitation, the cooling of the electron system can be expressed with two coupled differential equations, which describe the heat conduction of the electrons and the lattice:\cite{atxitia_evidence_2010}
\begin{equation}
    C_\text{e}\frac{\mathrm{d}T_\text{e}}{\mathrm{d}t} = -G_\text{el}(T_\text{e}-T_\text{l})+P(t)-C_\text{e}\frac{(T_\text{e}-T_\text{amb})}{\tau_\text{th}}
\label{eq:TTM1}
\end{equation}
\begin{equation}
    C_\text{l}\frac{\mathrm{d}T_\text{l}}{\mathrm{d}t} = -G_\text{el}(T_\text{e}-T_\text{l}) %\text{\qquad\cite{atxitia_evidence_2010}.}
\label{eq:TTM2}
\end{equation}
In these equations, $C_\text{e}$ and $C_\text{l}$ are the specific heats of the electrons and the lattice, respectively, $G_\text{el}$ is an electron-phonon coupling constant which connects the two equations and determines the rate of the energy exchange between the electrons and the lattice. $T_\text{amb}$ is the ambient room temperature and $\tau_\text{th}$ is the heat diffusion time. The first term in equations~\eqref{eq:TTM1} and~\eqref{eq:TTM2} describes the coupling between the two systems. In equation~\eqref{eq:TTM1}, the second term $P(t)$ represents the excitation source which is absorbed in the metal layer and excites the electron system.  The last term describes the heat diffusion in the metal layer.\\
\begin{figure}[t]
    \centering
    \includegraphics[width=0.9\columnwidth]{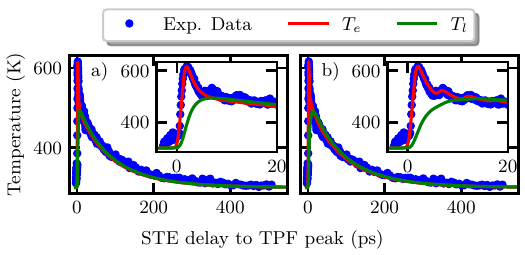}
    \caption{Decay of the maximum of the terahertz probe pulse induced peak in the STE terahertz waveform as shown in Figure~4a with additional fit. The inset shows the first 20\,ps of the decay. The fit is based on the two-temperature model and the electron temperature (red) as well as the lattice temperature (green) are displayed. In (a), the excitation source is approximated as a Gaussian pulse. Here, the fit is not in good agreement with the oscillations in the measurement. In (b), the excitation source is approximated as a sum of Gaussian pulses to account for multiple reflections in the sapphire substrate.}
    \label{fig:TTM_fit}
\end{figure}

To solve these equations, we make several simplifying assumptions. Normally, the electron specific heat is strongly temperature dependent. Here, we use the simplification $C_e = \gamma_e T_e$ with $\gamma_e=3\times 10^3\,\mathrm{Jm^{-3}K^{-1}}$, as proposed in Ref.~\citenum{atxitia_evidence_2010}. The temporal intensity of the excitation (pump) pulse is approximated as a Gaussian, and the absorbance of the laser pulse in the metal layer is approximated~\cite{kolejak_maximizing_2024} as 32\,\% and assumed to be uniform over the illuminated area and depth in the metal layer. It is assumed that both the maximum lattice temperature and the equilibrium between electron and lattice temperatures are reached when the fast decay in Figure~\ref{fig:waveform_decay_zoom} ends and the slow decay starts, because the coupling between the systems vanishes as the temperatures approach each other~\cite{agarwal_ultrafast_2021}. We approximate the surface total specific heat of the STE layer as $C_{\text{surface}} = 1.6\times 10^{-6}\,\mathrm{Jcm^{-2}K^{-1}}$ based on the thickness of the STE layer $d=\SI{6}{nm}$ and an estimated specific heat capacity of the STE of $C_{\text{specific}} = 2.7\times 10^{6}\,\mathrm{Jm^{-3}K^{-1}}$ based on the literature values for the specific heat capacity of platinum~\cite{yokokawa_laser-flash_1979} and tungsten~\cite{bronson_specific_1933}. By taking into account the repetition rate of the laser (\SI{1}{kHz}), and the average optical pump power (\SI{100}{mW}), as well as the pump spot area of \SI{0.1}{cm^2}, an absorbed fluence of $F = \SI{0.32}{\mathrm{ mJ/cm^{2}}}$ is derived. The maximum temperature of the STE with the electron and the lattice temperature in equilibrium is then estimated to be $T_\text{STE}^{\mathrm{max}} = T_\text{amb} + \Delta T = T_\text{amb} + F/C_{\text{surface}} \approx \SI{300}{K}+\SI{200}{K} = \SI{500}{K}$. This maximum temperature enables calibration of the scale in Figure~\ref{fig:TTM_fit}. It should be noted that the calculated absolute values have a large confidence interval, as the main parameters are approximated based on data of similar (but not identical) materials from literature.

The obtained fit, using the temperature-calibrated measurement data, is shown in Figure~\ref{fig:TTM_fit}. The lattice temperature is displayed in green. The calculated electron temperature and the measured data are in very good agreement. However, the inset in Figure~\ref{fig:TTM_fit}a shows that the oscillations due to the multiple reflections in the sapphire substrate are not captured by the model. From the fit to equations~\eqref{eq:TTM1} and~\eqref{eq:TTM2}, a heat diffusion time of $\tau_\text{th} = \SI{35}{ps}$  and an electron-phonon coupling constant of $G_\text{el} = 55\times 10^{16}\,\mathrm{ Wm^{-3}K^{-1}}$ are obtained, which are directly connected to the electron-phonon scattering rate and are of the same order of magnitude as reported in~\cite{atxitia_evidence_2010, agarwal_ultrafast_2021, beaurepaire_ultrafast_1996} for various comparable material systems.

Extending this model by modifying the excitation source $P(t)$ to a sum of Gaussian pulses provides a solution to account for the multiple reflections in the sapphire substrate. The fit with this modified excitation source is shown in Figure~\ref{fig:TTM_fit}b. Now, also the oscillations are in very good agreement with the fit. The small deviation of the measured data from the fit in the slow decay from \SI{20}{ps} to \SI{600}{ps} may be due to the excluded coupling term of the electronic system of the STE layer and the phononic system in the substrate~\cite{agarwal_ultrafast_2021}. 

From our results, we find that the electron system relaxes to its initial state about 500\,ps after excitation. Electronic and lattice temperatures in TTM studies typically equilibrate on a picosecond timescale, usually within tens of ps. Such a long decay points towards nonthermal effects or substrate coupling, which are normally neglected in the model. However, Radue \textit{et al.} have previously demonstrated that the thermoreflectance signal of thin gold films on sapphire substrates continues to evolve for hundreds of picoseconds. They attributed this extended decay to lattice thermal diffusion and interfacial thermal resistance, particularly between the metal film and the substrate.\cite{radue2018hot} The 500\,ps relaxation we observe suggests a threshold for the fluence-dominated damage process due to accumulated heat in the STE layer of around 2\,GHz. In a previous study~\cite{paries_optical_2024}, the optical damage process was connected to heat accumulation in the STE layer with a laser repetition-rate threshold of \SI{10}{MHz} for a fluence of $F = \SI{0.32}{\mathrm{mJ/cm^{2}}}$. This is two orders of magnitude lower than what we found here. 
The discrepancy can most likely be explained by the different material systems used in the study.
Taylor \textit{et al.} showed that the heat accumulation due to high repetition rate femtosecond lasers can be described by combining the TTM and a classical heat conduction equation \cite{taylor_integrating_2018, taylor_optimization_2016}. For the latter, the thermal conductivity $\lambda_\text{T}$ is a crucial material parameter. Comparing the previous study~\cite{paries_optical_2024} with the results here, the thermal conductivity in the fiber-tip STE glass substrate ($\lambda_\text{T, glass}\approx \SI{1}{W/mK}$) \cite{simoncelli_thermal_2023} is a factor of 30 weaker than the sapphire substrate used here ($\lambda_\text{T, sapphire}\approx \SI{30}{W/mK}$)\cite{pishchik_properties_2009}. This leads to a slower heat transfer away from the pump spot. Additionally, the fiber-tip STE used in the previous study provided a much smaller surface area of the STE (\SI{0.05}{cm^2} compared to \SI{1}{cm^2} used in the experiments presented here), which could lead to a saturated heat transfer to the surrounding area over time. This may explain the orders-of-magnitude higher repetition rate threshold in the current -- very localized -- experiment but can be investigated further. Recent studies have focused on optimizing  STEs for higher terahertz output power.  Limited heat transfer in the thin metallic STE layer was identified as one way to  improve the optical damage threshold of the STE, which impacts the terahertz emissivity~\cite{vogel_average_2022,vaitsi_rotating_2024, gandubert_spatial_2024}.

\section*{Conclusion}
In this work we present a new method of measuring the electron temperature dynamics in STEs using a terahertz probe pulse.
Our experiments show that optical-pump terahertz-probe measurements can be used to investigate the electron temperature dynamics and coupling mechanisms in STE structures. We find here that the electron system reaches a temperature of several 100\,K after excitation. However, the maximum electron temperature can only be estimated due to the time resolution of the terahertz probe pulse.
This method offers the distinct advantage that the electron temperature dynamics of the STE are not disturbed by the terahertz probe pulse, as it could easily be the case when using an optical probe pulse.
The current method enables for example the study of the optical damage process described in Ref.~\citenum{paries_optical_2024}, which is connected to heat accumulation in the STE layer. We find here that the electron system relaxes to its initial state after excitation after about \SI{500}{ps}. This suggests a threshold for the fluence-dominated damage process at around \SI{2}{GHz} for tri-layer STEs on sapphire, much higher than previously found~\cite{paries_optical_2024}. This difference is probably due to saturated heat transfer to the surrounding media. Further optical-pump terahertz-probe studies on the temperature relaxation could provide more quantitative insight on the thermodynamical processes causing heat accumulation and ultimately STE damage. A detailed investigation of the influence of substrate properties and pump spot size on the relaxation dynamics can offer further insight into the dominant cooling mechanisms in STEs. Such measurements, while beyond the scope of this work, represent a promising direction for future studies. Such studies may help to improve the thermal management of STEs, resulting in more efficient terahertz emitters. Furthermore, the presented method provides a minimally invasive probing approach that may be also interesting in future applications involving ultrafast dynamics, fragile materials, or resonant optical
transitions.


\subsection*{Funding}
We wish to acknowledge funding from NSF ECCS-2300152, Deutsche Forschungsgemein-
schaft (DFG, German Research Foundation) - TRR 173 - 268565370 (SPIN+X, project B11), and
a Study Abroad Studentship of the German Academic Scholarship Foundation (Studienstiftung
des deutschen Volkes).
%\nocite{*}
%\bibliography{aipsamp}% Produces the bibliography via BibTeX.

\begin{mcitethebibliography}{43}
\providecommand*\natexlab[1]{#1}
\providecommand*\mciteSetBstSublistMode[1]{}
\providecommand*\mciteSetBstMaxWidthForm[2]{}
\providecommand*\mciteBstWouldAddEndPuncttrue
  {\def\EndOfBibitem{\unskip.}}
\providecommand*\mciteBstWouldAddEndPunctfalse
  {\let\EndOfBibitem\relax}
\providecommand*\mciteSetBstMidEndSepPunct[3]{}
\providecommand*\mciteSetBstSublistLabelBeginEnd[3]{}
\providecommand*\EndOfBibitem{}
\mciteSetBstSublistMode{f}
\mciteSetBstMaxWidthForm{subitem}{(\alph{mcitesubitemcount})}
\mciteSetBstSublistLabelBeginEnd
  {\mcitemaxwidthsubitemform\space}
  {\relax}
  {\relax}

\bibitem[Kampfrath \latin{et~al.}(2013)Kampfrath, Battiato, Maldonado, Eilers, Nötzold, Mährlein, Zbarsky, Freimuth, Mokrousov, Blügel, Wolf, Radu, Oppeneer, and Münzenberg]{kampfrath_terahertz_2013}
Kampfrath,~T.; Battiato,~M.; Maldonado,~P.; Eilers,~G.; Nötzold,~J.; Mährlein,~S.; Zbarsky,~V.; Freimuth,~F.; Mokrousov,~Y.; Blügel,~S.; Wolf,~M.; Radu,~I.; Oppeneer,~P.~M.; Münzenberg,~M. Terahertz spin current pulses controlled by magnetic heterostructures. \emph{Nature Nanotechnology} \textbf{2013}, \emph{8}, 256--260\relax
\mciteBstWouldAddEndPuncttrue
\mciteSetBstMidEndSepPunct{\mcitedefaultmidpunct}
{\mcitedefaultendpunct}{\mcitedefaultseppunct}\relax
\EndOfBibitem
\bibitem[Seifert \latin{et~al.}(2022)Seifert, Cheng, Wei, Kampfrath, and Qi]{seifert_spintronic_2022}
Seifert,~T.~S.; Cheng,~L.; Wei,~Z.; Kampfrath,~T.; Qi,~J. Spintronic sources of ultrashort terahertz electromagnetic pulses. \emph{Applied Physics Letters} \textbf{2022}, \emph{120}, 180401\relax
\mciteBstWouldAddEndPuncttrue
\mciteSetBstMidEndSepPunct{\mcitedefaultmidpunct}
{\mcitedefaultendpunct}{\mcitedefaultseppunct}\relax
\EndOfBibitem
\bibitem[Torosyan \latin{et~al.}(2018)Torosyan, Keller, Scheuer, Beigang, and Papaioannou]{torosyan_optimized_2018}
Torosyan,~G.; Keller,~S.; Scheuer,~L.; Beigang,~R.; Papaioannou,~E.~T. Optimized {Spintronic} {Terahertz} {Emitters} {Based} on {Epitaxial} {Grown} {Fe}/{Pt} {Layer} {Structures}. \emph{Scientific Reports} \textbf{2018}, \emph{8}, 1311\relax
\mciteBstWouldAddEndPuncttrue
\mciteSetBstMidEndSepPunct{\mcitedefaultmidpunct}
{\mcitedefaultendpunct}{\mcitedefaultseppunct}\relax
\EndOfBibitem
\bibitem[Ilyakov \latin{et~al.}(2022)Ilyakov, Agarwal, Deinert, Liu, Yaroslavtsev, Foglia, Kurdi, Mincigrucci, Principi, Jakob, Kläui, Seifert, Kampfrath, Kovalev, Carley, Scherz, and Gensch]{ilyakov_terahertz-wave_2022}
Ilyakov,~I. \latin{et~al.}  Terahertz-wave decoding of femtosecond extreme-ultraviolet light pulses. \emph{Optica} \textbf{2022}, \emph{9}, 545\relax
\mciteBstWouldAddEndPuncttrue
\mciteSetBstMidEndSepPunct{\mcitedefaultmidpunct}
{\mcitedefaultendpunct}{\mcitedefaultseppunct}\relax
\EndOfBibitem
\bibitem[Papaioannou \latin{et~al.}(2018)Papaioannou, Torosyan, Keller, Scheuer, Battiato, Mag-Usara, L'huillier, Tani, and Beigang]{papaioannou_efficient_2018}
Papaioannou,~E.~T.; Torosyan,~G.; Keller,~S.; Scheuer,~L.; Battiato,~M.; Mag-Usara,~V.~K.; L'huillier,~J.; Tani,~M.; Beigang,~R. Efficient {Terahertz} {Generation} {Using} {Fe}/{Pt} {Spintronic} {Emitters} {Pumped} at {Different} {Wavelengths}. \emph{IEEE Transactions on Magnetics} \textbf{2018}, \emph{54}, 1--5\relax
\mciteBstWouldAddEndPuncttrue
\mciteSetBstMidEndSepPunct{\mcitedefaultmidpunct}
{\mcitedefaultendpunct}{\mcitedefaultseppunct}\relax
\EndOfBibitem
\bibitem[Nandi \latin{et~al.}(2019)Nandi, Abdelaziz, Jaiswal, Jakob, Gueckstock, Rouzegar, Seifert, Kläui, Kampfrath, and Preu]{nandi_antenna-coupled_2019}
Nandi,~U.; Abdelaziz,~M.~S.; Jaiswal,~S.; Jakob,~G.; Gueckstock,~O.; Rouzegar,~S.~M.; Seifert,~T.~S.; Kläui,~M.; Kampfrath,~T.; Preu,~S. Antenna-coupled spintronic terahertz emitters driven by a 1550 nm femtosecond laser oscillator. \emph{Applied Physics Letters} \textbf{2019}, \emph{115}, 022405\relax
\mciteBstWouldAddEndPuncttrue
\mciteSetBstMidEndSepPunct{\mcitedefaultmidpunct}
{\mcitedefaultendpunct}{\mcitedefaultseppunct}\relax
\EndOfBibitem
\bibitem[Ilyakov \latin{et~al.}(2023)Ilyakov, Brataas, De~Oliveira, Ponomaryov, Deinert, Hellwig, Faßbender, Lindner, Salikhov, and Kovalev]{ilyakov_efficient_2023}
Ilyakov,~I.; Brataas,~A.; De~Oliveira,~T. V. A.~G.; Ponomaryov,~A.; Deinert,~J.-C.; Hellwig,~O.; Faßbender,~J.; Lindner,~J.; Salikhov,~R.; Kovalev,~S. Efficient ultrafast field-driven spin current generation for spintronic terahertz frequency conversion. \emph{Nature Communications} \textbf{2023}, \emph{14}, 7010\relax
\mciteBstWouldAddEndPuncttrue
\mciteSetBstMidEndSepPunct{\mcitedefaultmidpunct}
{\mcitedefaultendpunct}{\mcitedefaultseppunct}\relax
\EndOfBibitem
\bibitem[Seifert \latin{et~al.}(2016)Seifert, Jaiswal, Martens, Hannegan, Braun, Maldonado, Freimuth, Kronenberg, Henrizi, Radu, Beaurepaire, Mokrousov, Oppeneer, Jourdan, Jakob, Turchinovich, Hayden, Wolf, Münzenberg, Kläui, and Kampfrath]{seifert_efficient_2016}
Seifert,~T. \latin{et~al.}  Efficient metallic spintronic emitters of ultrabroadband terahertz radiation. \emph{Nature Photonics} \textbf{2016}, \emph{10}, 483--488\relax
\mciteBstWouldAddEndPuncttrue
\mciteSetBstMidEndSepPunct{\mcitedefaultmidpunct}
{\mcitedefaultendpunct}{\mcitedefaultseppunct}\relax
\EndOfBibitem
\bibitem[Paries \latin{et~al.}(2024)Paries, Selz, Santos, Lampin, Koleják, Lezier, Troadec, Tiercelin, Vanwolleghem, Addda, Kampfrath, Seifert, Freymann, and Molter]{paries_optical_2024}
Paries,~F.; Selz,~F.; Santos,~C.~N.; Lampin,~J.-F.; Koleják,~P.; Lezier,~G.; Troadec,~D.; Tiercelin,~N.; Vanwolleghem,~M.; Addda,~A.; Kampfrath,~T.; Seifert,~T.~S.; Freymann,~G.~V.; Molter,~D. Optical damage thresholds of single-mode fiber-tip spintronic terahertz emitters. \emph{Optics Express} \textbf{2024}, \emph{32}, 24826\relax
\mciteBstWouldAddEndPuncttrue
\mciteSetBstMidEndSepPunct{\mcitedefaultmidpunct}
{\mcitedefaultendpunct}{\mcitedefaultseppunct}\relax
\EndOfBibitem
\bibitem[Agarwal \latin{et~al.}(2021)Agarwal, Medwal, Kumar, Asada, Fukuma, Rawat, Battiato, and Singh]{agarwal_ultrafast_2021}
Agarwal,~P.; Medwal,~R.; Kumar,~A.; Asada,~H.; Fukuma,~Y.; Rawat,~R.~S.; Battiato,~M.; Singh,~R. Ultrafast {Photo}‐{Thermal} {Switching} of {Terahertz} {Spin} {Currents}. \emph{Advanced Functional Materials} \textbf{2021}, \emph{31}, 2010453\relax
\mciteBstWouldAddEndPuncttrue
\mciteSetBstMidEndSepPunct{\mcitedefaultmidpunct}
{\mcitedefaultendpunct}{\mcitedefaultseppunct}\relax
\EndOfBibitem
\bibitem[Vogel \latin{et~al.}(2022)Vogel, Omar, Mansourzadeh, Wulf, Sabanés, Müller, Seifert, Weigel, Jakob, Kläui, Pupeza, Kampfrath, and Saraceno]{vogel_average_2022}
Vogel,~T.; Omar,~A.; Mansourzadeh,~S.; Wulf,~F.; Sabanés,~N.~M.; Müller,~M.; Seifert,~T.~S.; Weigel,~A.; Jakob,~G.; Kläui,~M.; Pupeza,~I.; Kampfrath,~T.; Saraceno,~C.~J. Average power scaling of {THz} spintronic emitters efficiently cooled in reflection geometry. \emph{Optics Express} \textbf{2022}, \emph{30}, 20451\relax
\mciteBstWouldAddEndPuncttrue
\mciteSetBstMidEndSepPunct{\mcitedefaultmidpunct}
{\mcitedefaultendpunct}{\mcitedefaultseppunct}\relax
\EndOfBibitem
\bibitem[Vaitsi \latin{et~al.}(2024)Vaitsi, Sleziona, Parra~López, Behovits, Schulz, Martín~Sabanés, Kampfrath, Wolf, Seifert, and Müller]{vaitsi_rotating_2024}
Vaitsi,~A.; Sleziona,~V.; Parra~López,~L.~E.; Behovits,~Y.; Schulz,~F.; Martín~Sabanés,~N.; Kampfrath,~T.; Wolf,~M.; Seifert,~T.~S.; Müller,~M. Rotating spintronic terahertz emitter optimized for microjoule pump-pulse energies and megahertz repetition rates. \emph{Applied Physics Letters} \textbf{2024}, \emph{125}, 071107\relax
\mciteBstWouldAddEndPuncttrue
\mciteSetBstMidEndSepPunct{\mcitedefaultmidpunct}
{\mcitedefaultendpunct}{\mcitedefaultseppunct}\relax
\EndOfBibitem
\bibitem[Gandubert \latin{et~al.}(2024)Gandubert, Nkeck, Ropagnol, Morris, and Blanchard]{gandubert_spatial_2024}
Gandubert,~G.; Nkeck,~J.~E.; Ropagnol,~X.; Morris,~D.; Blanchard,~F. Spatial and temporal thermal management of a spintronic terahertz emitter. \emph{Applied Physics Express} \textbf{2024}, \emph{17}, 083001\relax
\mciteBstWouldAddEndPuncttrue
\mciteSetBstMidEndSepPunct{\mcitedefaultmidpunct}
{\mcitedefaultendpunct}{\mcitedefaultseppunct}\relax
\EndOfBibitem
\bibitem[Fann \latin{et~al.}(1992)Fann, Storz, Tom, and Bokor]{fann_direct_1992}
Fann,~W.~S.; Storz,~R.; Tom,~H. W.~K.; Bokor,~J. Direct measurement of nonequilibrium electron-energy distributions in subpicosecond laser-heated gold films. \emph{Physical Review Letters} \textbf{1992}, \emph{68}, 2834--2837, Publisher: American Physical Society (APS)\relax
\mciteBstWouldAddEndPuncttrue
\mciteSetBstMidEndSepPunct{\mcitedefaultmidpunct}
{\mcitedefaultendpunct}{\mcitedefaultseppunct}\relax
\EndOfBibitem
\bibitem[Volkov \latin{et~al.}(1976)Volkov, Palatnik, and Pugachev]{volkov_investigation_1976}
Volkov,~Y.~A.; Palatnik,~L.~S.; Pugachev,~A.~T. Investigation of the thermal properties of thin aluminum films. \emph{Soviet Journal of Experimental and Theoretical Physics} \textbf{1976}, \emph{43}, 1171\relax
\mciteBstWouldAddEndPuncttrue
\mciteSetBstMidEndSepPunct{\mcitedefaultmidpunct}
{\mcitedefaultendpunct}{\mcitedefaultseppunct}\relax
\EndOfBibitem
\bibitem[Jollans \latin{et~al.}(2020)Jollans, Caldarola, Sivan, and Orrit]{jollans_effective_2020}
Jollans,~T.; Caldarola,~M.; Sivan,~Y.; Orrit,~M. Effective {Electron} {Temperature} {Measurement} {Using} {Time}-{Resolved} {Anti}-{Stokes} {Photoluminescence}. \emph{The Journal of Physical Chemistry A} \textbf{2020}, \emph{124}, 6968--6976, Publisher: American Chemical Society (ACS)\relax
\mciteBstWouldAddEndPuncttrue
\mciteSetBstMidEndSepPunct{\mcitedefaultmidpunct}
{\mcitedefaultendpunct}{\mcitedefaultseppunct}\relax
\EndOfBibitem
\bibitem[Hayashi \latin{et~al.}(2022)Hayashi, Iwasaki, Vasa, and Yamanouchi]{hayashi_determination_2022}
Hayashi,~R.; Iwasaki,~A.; Vasa,~P.; Yamanouchi,~K. Determination of electron and phonon temperatures in gold thin film irradiated with an ultrashort laser pulse. \emph{AIP Advances} \textbf{2022}, \emph{12}, 095207\relax
\mciteBstWouldAddEndPuncttrue
\mciteSetBstMidEndSepPunct{\mcitedefaultmidpunct}
{\mcitedefaultendpunct}{\mcitedefaultseppunct}\relax
\EndOfBibitem
\bibitem[Hohlfeld \latin{et~al.}(2000)Hohlfeld, Wellershoff, G{\"u}dde, Conrad, J{\"a}hnke, and Matthias]{hohlfeld2000electron}
Hohlfeld,~J.; Wellershoff,~S.-S.; G{\"u}dde,~J.; Conrad,~U.; J{\"a}hnke,~V.; Matthias,~E. Electron and lattice dynamics following optical excitation of metals. \emph{Chemical physics} \textbf{2000}, \emph{251}, 237--258\relax
\mciteBstWouldAddEndPuncttrue
\mciteSetBstMidEndSepPunct{\mcitedefaultmidpunct}
{\mcitedefaultendpunct}{\mcitedefaultseppunct}\relax
\EndOfBibitem
\bibitem[Beard \latin{et~al.}(2002)Beard, Turner, and Schmuttenmaer]{beard_terahertz_2002}
Beard,~M.~C.; Turner,~G.~M.; Schmuttenmaer,~C.~A. Terahertz {Spectroscopy}. \emph{The Journal of Physical Chemistry B} \textbf{2002}, \emph{106}, 7146--7159\relax
\mciteBstWouldAddEndPuncttrue
\mciteSetBstMidEndSepPunct{\mcitedefaultmidpunct}
{\mcitedefaultendpunct}{\mcitedefaultseppunct}\relax
\EndOfBibitem
\bibitem[Boyd \latin{et~al.}(2008)Boyd, Gaeta, and Giese]{boyd2008nonlinear}
Boyd,~R.~W.; Gaeta,~A.~L.; Giese,~E. \emph{Springer Handbook of Atomic, Molecular, and Optical Physics}; Springer, 2008; pp 1097--1110\relax
\mciteBstWouldAddEndPuncttrue
\mciteSetBstMidEndSepPunct{\mcitedefaultmidpunct}
{\mcitedefaultendpunct}{\mcitedefaultseppunct}\relax
\EndOfBibitem
\bibitem[Selz \latin{et~al.}(2025)Selz, K\"{o}lbel, Paries, von Freymann, Molter, and Mittleman]{Selz_25}
Selz,~F.; K\"{o}lbel,~J.; Paries,~F.; von Freymann,~G.; Molter,~D.; Mittleman,~D.~M. Terahertz-induced nonlinear response in {ZnTe}. \emph{Opt. Express} \textbf{2025}, \emph{33}, 9575--9586\relax
\mciteBstWouldAddEndPuncttrue
\mciteSetBstMidEndSepPunct{\mcitedefaultmidpunct}
{\mcitedefaultendpunct}{\mcitedefaultseppunct}\relax
\EndOfBibitem
\bibitem[Hebling \latin{et~al.}(2008)Hebling, Yeh, Hoffmann, Bartal, and Nelson]{Hebling_08}
Hebling,~J.; Yeh,~K.-L.; Hoffmann,~M.~C.; Bartal,~B.; Nelson,~K.~A. Generation of high-power terahertz pulses by tilted-pulse-front excitation and their application possibilities. \emph{J. Opt. Soc. Am. B} \textbf{2008}, \emph{25}, B6--B19\relax
\mciteBstWouldAddEndPuncttrue
\mciteSetBstMidEndSepPunct{\mcitedefaultmidpunct}
{\mcitedefaultendpunct}{\mcitedefaultseppunct}\relax
\EndOfBibitem
\bibitem[Koch \latin{et~al.}(2023)Koch, Mittleman, Ornik, and Castro-Camus]{koch_terahertz_2023}
Koch,~M.; Mittleman,~D.~M.; Ornik,~J.; Castro-Camus,~E. Terahertz time-domain spectroscopy. \emph{Nature Reviews Methods Primers} \textbf{2023}, \emph{3}, 48\relax
\mciteBstWouldAddEndPuncttrue
\mciteSetBstMidEndSepPunct{\mcitedefaultmidpunct}
{\mcitedefaultendpunct}{\mcitedefaultseppunct}\relax
\EndOfBibitem
\bibitem[Paries \latin{et~al.}(2023)Paries, Tiercelin, Lezier, Vanwolleghem, Selz, Syskaki, Kammerbauer, Jakob, Jourdan, Kläui, Kaspar, Kampfrath, Seifert, Von~Freymann, and Molter]{paries_fiber-tip_2023}
Paries,~F.; Tiercelin,~N.; Lezier,~G.; Vanwolleghem,~M.; Selz,~F.; Syskaki,~M.-A.; Kammerbauer,~F.; Jakob,~G.; Jourdan,~M.; Kläui,~M.; Kaspar,~Z.; Kampfrath,~T.; Seifert,~T.~S.; Von~Freymann,~G.; Molter,~D. Fiber-tip spintronic terahertz emitters. \emph{Optics Express} \textbf{2023}, \emph{31}, 30884\relax
\mciteBstWouldAddEndPuncttrue
\mciteSetBstMidEndSepPunct{\mcitedefaultmidpunct}
{\mcitedefaultendpunct}{\mcitedefaultseppunct}\relax
\EndOfBibitem
\bibitem[Planken \latin{et~al.}(2001)Planken, Nienhuys, Bakker, and Wenckebach]{planken2001measurement}
Planken,~P.~C.; Nienhuys,~H.-K.; Bakker,~H.~J.; Wenckebach,~T. Measurement and calculation of the orientation dependence of terahertz pulse detection in {ZnTe}. \emph{JOSA B} \textbf{2001}, \emph{18}, 313--317\relax
\mciteBstWouldAddEndPuncttrue
\mciteSetBstMidEndSepPunct{\mcitedefaultmidpunct}
{\mcitedefaultendpunct}{\mcitedefaultseppunct}\relax
\EndOfBibitem
\bibitem[Wu \latin{et~al.}(2021)Wu, Yaw~Ameyaw, Doty, and Jungfleisch]{wu2021principles}
Wu,~W.; Yaw~Ameyaw,~C.; Doty,~M.~F.; Jungfleisch,~M.~B. Principles of spintronic {THz} emitters. \emph{Journal of Applied Physics} \textbf{2021}, \emph{130}\relax
\mciteBstWouldAddEndPuncttrue
\mciteSetBstMidEndSepPunct{\mcitedefaultmidpunct}
{\mcitedefaultendpunct}{\mcitedefaultseppunct}\relax
\EndOfBibitem
\bibitem[Thomas \latin{et~al.}(1998)Thomas, Andersson, Sova, and Joseph]{thomas_frequency_1998}
Thomas,~M.~E.; Andersson,~S.~K.; Sova,~R.~M.; Joseph,~R.~I. Frequency and temperature dependence of the refractive index of sapphire. \emph{Infrared Physics \& Technology} \textbf{1998}, \emph{39}, 235--249\relax
\mciteBstWouldAddEndPuncttrue
\mciteSetBstMidEndSepPunct{\mcitedefaultmidpunct}
{\mcitedefaultendpunct}{\mcitedefaultseppunct}\relax
\EndOfBibitem
\bibitem[Bigot \latin{et~al.}(2009)Bigot, Vomir, and Beaurepaire]{bigot_coherent_2009}
Bigot,~J.-Y.; Vomir,~M.; Beaurepaire,~E. Coherent ultrafast magnetism induced by femtosecond laser pulses. \emph{Nature Physics} \textbf{2009}, \emph{5}, 515--520\relax
\mciteBstWouldAddEndPuncttrue
\mciteSetBstMidEndSepPunct{\mcitedefaultmidpunct}
{\mcitedefaultendpunct}{\mcitedefaultseppunct}\relax
\EndOfBibitem
\bibitem[Guo \latin{et~al.}(2012)Guo, Hodson, Fisher, and Xu]{guo_heat_2011}
Guo,~L.; Hodson,~S.~L.; Fisher,~T.~S.; Xu,~X. Heat {Transfer} {Across} {Metal}-{Dielectric} {Interfaces} {During} {Ultrafast}-{Laser} {Heating}. \emph{Journal of Heat Transfer} \textbf{2012}, \emph{134}, 042402\relax
\mciteBstWouldAddEndPuncttrue
\mciteSetBstMidEndSepPunct{\mcitedefaultmidpunct}
{\mcitedefaultendpunct}{\mcitedefaultseppunct}\relax
\EndOfBibitem
\bibitem[Anisimov \latin{et~al.}(1974)Anisimov, Kapeliovich, and Perelman]{anisimov_electron_1974}
Anisimov,~S.~I.; Kapeliovich,~B.~L.; Perelman,~T.~L. Electron emission from metal surfaces exposed to ultrashort laser pulses. \emph{Soviet Journal of Experimental and Theoretical Physics} \textbf{1974}, \emph{39}, 375--377\relax
\mciteBstWouldAddEndPuncttrue
\mciteSetBstMidEndSepPunct{\mcitedefaultmidpunct}
{\mcitedefaultendpunct}{\mcitedefaultseppunct}\relax
\EndOfBibitem
\bibitem[Kaganov \latin{et~al.}(1957)Kaganov, Lifshitz, and Tanatarov]{kaganov_relaxation_1957}
Kaganov,~M.~L.; Lifshitz,~I.~M.; Tanatarov,~L.~V. Relaxation between electrons and the crystalline lattice. \emph{Soviet Journal of Experimental and Theoretical Physics} \textbf{1957}, \emph{4}, 173--178\relax
\mciteBstWouldAddEndPuncttrue
\mciteSetBstMidEndSepPunct{\mcitedefaultmidpunct}
{\mcitedefaultendpunct}{\mcitedefaultseppunct}\relax
\EndOfBibitem
\bibitem[Atxitia \latin{et~al.}(2010)Atxitia, Chubykalo-Fesenko, Walowski, Mann, and Münzenberg]{atxitia_evidence_2010}
Atxitia,~U.; Chubykalo-Fesenko,~O.; Walowski,~J.; Mann,~A.; Münzenberg,~M. Evidence for thermal mechanisms in laser-induced femtosecond spin dynamics. \emph{Physical Review B} \textbf{2010}, \emph{81}, 174401\relax
\mciteBstWouldAddEndPuncttrue
\mciteSetBstMidEndSepPunct{\mcitedefaultmidpunct}
{\mcitedefaultendpunct}{\mcitedefaultseppunct}\relax
\EndOfBibitem
\bibitem[Giri and Hopkins(2020)Giri, and Hopkins]{giri2020review}
Giri,~A.; Hopkins,~P.~E. A review of experimental and computational advances in thermal boundary conductance and nanoscale thermal transport across solid interfaces. \emph{Advanced Functional Materials} \textbf{2020}, \emph{30}, 1903857\relax
\mciteBstWouldAddEndPuncttrue
\mciteSetBstMidEndSepPunct{\mcitedefaultmidpunct}
{\mcitedefaultendpunct}{\mcitedefaultseppunct}\relax
\EndOfBibitem
\bibitem[Koleják \latin{et~al.}(2024)Koleják, Lezier, Vala, Mathmann, Halagačka, Gelnárová, Dusch, Lampin, Tiercelin, Postava, and Vanwolleghem]{kolejak_maximizing_2024}
Koleják,~P.; Lezier,~G.; Vala,~D.; Mathmann,~B.; Halagačka,~L.; Gelnárová,~Z.; Dusch,~Y.; Lampin,~J.; Tiercelin,~N.; Postava,~K.; Vanwolleghem,~M. Maximizing the {Electromagnetic} {Efficiency} of {Spintronic} {Terahertz} {Emitters}. \emph{Advanced Photonics Research} \textbf{2024}, \emph{5}, 2400064\relax
\mciteBstWouldAddEndPuncttrue
\mciteSetBstMidEndSepPunct{\mcitedefaultmidpunct}
{\mcitedefaultendpunct}{\mcitedefaultseppunct}\relax
\EndOfBibitem
\bibitem[Yokokawa and Takahashi(1979)Yokokawa, and Takahashi]{yokokawa_laser-flash_1979}
Yokokawa,~H.; Takahashi,~Y. Laser-flash calorimetry {II}. {Heat} capacity of platinum from 80 to 1000 {K} and its revised thermodynamic functions. \emph{The Journal of Chemical Thermodynamics} \textbf{1979}, \emph{11}, 411--420\relax
\mciteBstWouldAddEndPuncttrue
\mciteSetBstMidEndSepPunct{\mcitedefaultmidpunct}
{\mcitedefaultendpunct}{\mcitedefaultseppunct}\relax
\EndOfBibitem
\bibitem[Bronson \latin{et~al.}(1933)Bronson, Chisholm, and Dockerty]{bronson_specific_1933}
Bronson,~H.~L.; Chisholm,~H.~M.; Dockerty,~S.~M. {On} {the} {specific} {heats} {of} {tungsten}, {molybden}, {and} {copper}. \emph{Canadian Journal of Research} \textbf{1933}, \emph{8}, 282--303\relax
\mciteBstWouldAddEndPuncttrue
\mciteSetBstMidEndSepPunct{\mcitedefaultmidpunct}
{\mcitedefaultendpunct}{\mcitedefaultseppunct}\relax
\EndOfBibitem
\bibitem[Beaurepaire \latin{et~al.}(1996)Beaurepaire, Merle, Daunois, and Bigot]{beaurepaire_ultrafast_1996}
Beaurepaire,~E.; Merle,~J.-C.; Daunois,~A.; Bigot,~J.-Y. Ultrafast {Spin} {Dynamics} in {Ferromagnetic} {Nickel}. \emph{Physical Review Letters} \textbf{1996}, \emph{76}, 4250--4253\relax
\mciteBstWouldAddEndPuncttrue
\mciteSetBstMidEndSepPunct{\mcitedefaultmidpunct}
{\mcitedefaultendpunct}{\mcitedefaultseppunct}\relax
\EndOfBibitem
\bibitem[Radue \latin{et~al.}(2018)Radue, Tomko, Giri, Braun, Zhou, Prezhdo, Runnerstrom, Maria, and Hopkins]{radue2018hot}
Radue,~E.~L.; Tomko,~J.~A.; Giri,~A.; Braun,~J.~L.; Zhou,~X.; Prezhdo,~O.~V.; Runnerstrom,~E.~L.; Maria,~J.-P.; Hopkins,~P.~E. Hot electron thermoreflectance coefficient of gold during electron--phonon nonequilibrium. \emph{Acs Photonics} \textbf{2018}, \emph{5}, 4880--4887\relax
\mciteBstWouldAddEndPuncttrue
\mciteSetBstMidEndSepPunct{\mcitedefaultmidpunct}
{\mcitedefaultendpunct}{\mcitedefaultseppunct}\relax
\EndOfBibitem
\bibitem[Taylor \latin{et~al.}(2018)Taylor, Scott, and Qiao]{taylor_integrating_2018}
Taylor,~L.~L.; Scott,~R.~E.; Qiao,~J. Integrating two-temperature and classical heat accumulation models to predict femtosecond laser processing of silicon. \emph{Optical Materials Express} \textbf{2018}, \emph{8}, 648, Publisher: Optica Publishing Group\relax
\mciteBstWouldAddEndPuncttrue
\mciteSetBstMidEndSepPunct{\mcitedefaultmidpunct}
{\mcitedefaultendpunct}{\mcitedefaultseppunct}\relax
\EndOfBibitem
\bibitem[Taylor \latin{et~al.}(2016)Taylor, Qiao, and Qiao]{taylor_optimization_2016}
Taylor,~L.~L.; Qiao,~J.; Qiao,~J. Optimization of femtosecond laser processing of silicon via numerical modeling. \emph{Optical Materials Express} \textbf{2016}, \emph{6}, 2745, Publisher: Optica Publishing Group\relax
\mciteBstWouldAddEndPuncttrue
\mciteSetBstMidEndSepPunct{\mcitedefaultmidpunct}
{\mcitedefaultendpunct}{\mcitedefaultseppunct}\relax
\EndOfBibitem
\bibitem[Simoncelli \latin{et~al.}(2023)Simoncelli, Mauri, and Marzari]{simoncelli_thermal_2023}
Simoncelli,~M.; Mauri,~F.; Marzari,~N. Thermal conductivity of glasses: first-principles theory and applications. \emph{npj computational materials} \textbf{2023}, \emph{9}, 106\relax
\mciteBstWouldAddEndPuncttrue
\mciteSetBstMidEndSepPunct{\mcitedefaultmidpunct}
{\mcitedefaultendpunct}{\mcitedefaultseppunct}\relax
\EndOfBibitem
\bibitem[Dobrovinskaya \latin{et~al.}(2009)Dobrovinskaya, Lytvynov, and Pishchik]{pishchik_properties_2009}
Dobrovinskaya,~E.~R.; Lytvynov,~L.~A.; Pishchik,~V. \emph{Sapphire}; Springer US: Boston, MA, 2009; pp 55--176\relax
\mciteBstWouldAddEndPuncttrue
\mciteSetBstMidEndSepPunct{\mcitedefaultmidpunct}
{\mcitedefaultendpunct}{\mcitedefaultseppunct}\relax
\EndOfBibitem
\end{mcitethebibliography}

\providecommand{\latin}[1]{#1}
\makeatletter
\providecommand{\doi}
  {\begingroup\let\do\@makeother\dospecials
  \catcode`\{=1 \catcode`\}=2 \doi@aux}
\providecommand{\doi@aux}[1]{\endgroup\texttt{#1}}
\makeatother
\providecommand*\mcitethebibliography{\thebibliography}
\csname @ifundefined\endcsname{endmcitethebibliography}  {\let\endmcitethebibliography\endthebibliography}{}

\end{document}